# Optical vortices in nanophotonics

Chenhao Li[1], Stefan A. Maier[1,2], Haoran Ren[3]

[1] Chair in Hybrid Nanosystems, Nanoinstitute Munich, Faculty of Physics, Ludwig Maximilians-Universität München, München, 80539, Germany

[2] Department of Physics, Imperial College London, London SW7 2AZ, United Kingdom

[3] MQ Photonics Research Centre, Department of Physics and Astronomy, Macquarie University, Australia

Email: Haoran.Ren@mq.edu.au

**Abstract**

In the last two decades, optical vortices carried by twisted light wavefronts have attracted a great deal of interest, providing not only new physical insights into light-matter interactions, but also a transformative platform for boosting optical information capacity. Meanwhile, advances in nanoscience and nanotechnology lead to the emerging field of nanophotonics, offering an unprecedented level of light manipulation via nanostructured materials and devices. Many exciting ideas and concepts come up when optical vortices meet nanophotonic devices. Here, we provide a mini review on recent achievements made in nanophotonics for the generation and detection of optical vortices and some of their applications.

1. Introduction

With the rapid advancement of information technology, optical communication that allows an improved data transmission rate for long distances has become an essential part of our daily life. Developing high-capacity optical communications for free-space wireless and fiber-optic networks is of great importance for meeting the ever-increasing demands for high-speed internet, augmented reality, and the internet of things. Unlike electronic devices being able to address signals only in the temporal domain, optical devices can process information coded in both spatial and temporal domains, significantly improving the capacity of optical devices. In this context, optical multiplexing has been developed as a viable approach for increasing the capacity of optical communication systems, through encoding and decoding optical information channels onto different degrees of freedom, including amplitude, phase, polarization, wavelength, and the angular momentum of light.

Even though optical multiplexing opens the door to boost the capacity of optical communication systems, it generally uses bulky optical devices with large footprints to process the (de)multiplexed optical information. This has limited the scope of integrating high-capacity optical multiplexing systems for on-chip photonic and quantum information processing. Nanophotonics, a recently emerged field at the interface between nanotechnology and photonics, offers an unprecedented level of light manipulation via nanostructured materials and devices. In this context, nanophotonic multiplexing of light opens the possibility of miniaturizing high-capacity optical devices and systems for on-chip applications, such as multiprocessor system-on-chip[1-4] and integrated photonic quantum technologies[5-11] In the last decade, nanophotonic multiplexing of light has been demonstrated through nanostructured devices offering distinctive optical responses to the polarization[12-16], wavelength[14, 17-21], lifetime[22], and orbital angular momentum (OAM)[23, 24] of light. In particular, the OAM degree of freedom has recently attracted a great deal of interest, due to the fact that its physically unbounded set of orthogonal spatial modes holds strong promise to boost the multiplexing capacity.

## 2. Angular momentum of light

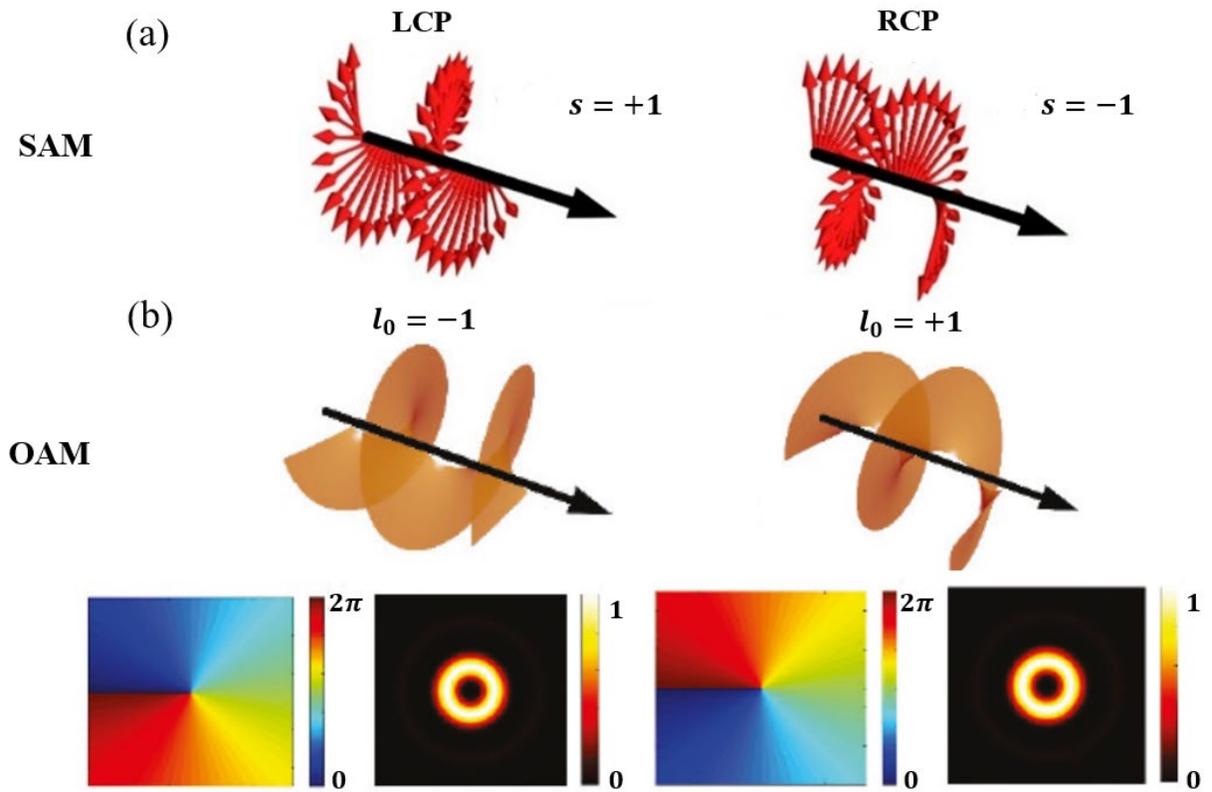

**Fig. 1. Schematics of the SAM and OAM of light.** (a) Left or right circular polarization light with SAM of 1 or -1 (b) Two kinds of light carrying different OAM states with -1, 1 as well as their phase and intensity distribution at bottom row, respectively.

The angular momentum is a property to describe the rotation of an object around an axis. When it applies to light, the angular momentum **J** can be defined as the cross-product between the position vector **r** and the linear momentum vector:

$$J = \varepsilon_0 \int \mathbf{r} \times (\mathbf{E} \times \mathbf{B}) \, d^3 r, \tag{1}$$

where $\varepsilon_0$ is the vacuum permittivity, **E** and **B** are the vectorial electric and magnetic fields of an electromagnetic wave. The angular momentum can be separated in the paraxial limit into two parts: the spin angular momentum (SAM) associated with circular polarization, and the OAM manifested by an optical vortex beam with a helical wavefront. For both paraxial and nonparaxial beams (e.g., a tightly focused beam and near-field optics), the total angular momentum:

$$J = s + l, \quad (2)$$

is a conserved quantity, where $s$ and $l$ are the quantum numbers of the SAM and OAM, respectively.

The discovery of the SAM of light can be traced back to Poynting[25] who suggested that a circularly polarized beam of wavelength $\lambda$ has an associated intrinsic angular momentum equivalent to $\lambda/2\pi$ multiplied by the linear momentum of the light beam. The linear momentum of light is given by $\hbar k$ per photon, where $k = 2\pi/\lambda$ is the wavevector of light. This leads to the conclusion that circularly polarized light carries SAM of $-\hbar$ or $+\hbar$ per photon. Beth suspended a half-wave plate and measured torque caused by passing circularly polarized light through the half-wave plate with changed handedness (e.g., from left- to right-handed or vice versa)[26]. The torque resulted from the handedness change of the circularly polarized light transferred $2\hbar$ of angular momentum to the plate, owing to the conservation of angular momentum during the light and wave plate interaction. Therefore, a circularly polarized optical beam carries the SAM with two possible spin states $\pm\hbar$, depending on its handedness (Fig. 1a).

In contrast to the intrinsic SAM related to the vector nature of light, light can also carry OAM having both intrinsic and extrinsic terms[27] and the latter of which is coordinate dependent. The intrinsic OAM, hereafter simply named as OAM, is carried by a helical wavefront that has a phase twist along the azimuthal direction in the transverse plane of an optical beam. In 1992, Allen et al. pioneered the discovery of the OAM in an optical vortex beam, with a helical phase distribution that mathematically satisfies exp $(il\varphi)$[28], where $l$ is the azimuthal mode index and $\varphi$ is the azimuthal angle of an OAM beam, respectively. Notably, the OAM degree of freedom has a theoretically unbounded set of eigen-states $\pm l\hbar$, where the integer number $l$ determines how strong the wavefront is azimuthally twisted (Fig. 1b). More strikingly, all OAM eigen-states are mutually orthogonal, opening the possibility of using OAM modes as independent information channels to carry optical information with negligible crosstalk. Therefore, in the last decade, OAM multiplexing has attracted a great deal of attention for high-capacity optical and quantum communications[29-31], and the OAM of light holds great promise to significantly improve the capacity of optical devices and systems.

It should be noted that there are some practical limitations in the use of higher order OAM vortex modes, although the OAM has a theoretically infinite number of orthogonal modes. Firstly, the number of cycles of phase shifts (from 0 to 2π) in the azimuthal direction determines the helical mode index of an OAM vortex mode, as such, it is challenging for practical optical devices (e.g., spatial light modulators) to implement higher-order OAM modes (e.g., $|l|$>500) that require rapid phase changes. Secondly, even though increasing the aperture size of optical devices could accommodate more rapid phase changes for higher-order OAM modes, the increased size prevents their devices from compact OAM generation and detection. Thirdly, higher-order OAM beams diverge faster than their lower-order counterparts and become more lossy in free-space propagation[32], setting up another practical limit to manipulate and detect higher-order OAM modes. However, it is worth mentioning that current optical devices can already implement a relatively large number of OAM modes ($|l|>100$) that are adequate for most of applications.

Even though an OAM-carrying optical vortex beam can be produced and detected via free-space optics based on either refractive or diffractive optical elements, such as cylindrical lenses[28], compute-generated holograms[33, 34], spiral phase plates[35], the footprint of the associated optical devices are too large to achieve photonic integration. In recent decades, nanophotonics has opened the door to device miniaturization, leading to a diverse range of compact and multifunctional optical devices harnessing the generation and detection of optical vortices, such as plasmonic and dielectric OAM metasurfaces[36-40], OAM vortex emitters[41, 42], vortex microlasers[43-45], plasmonic vortex devices[46-48], OAM photodetectors[49, 50], and on-chip OAM multiplexers[23, 24]. Here, we review recent achievements made in nanophotonics for the generation and detection of optical vortices and some of their applications.

3. **Nanophotonic generation of optical vortices**

Optical vortices carrying the OAM of light have played a vital role in both fundamental light-matter interactions and photonic applications. There is a continuing trend towards miniaturization of photonic devices for creating optical vortices. In the following section, we will review recent breakthroughs in nanophotonic generation of optical vortices and highlight some milestone works in the field. Based on their implementation principles, nanophotonic devices used for the generation of optical vortices are mainly divided into two categories: metasurfaces[36-40], and plasmonic and dielectric resonators[41-44].

3.1. **Metasurfaces-based generation of optical vortices**

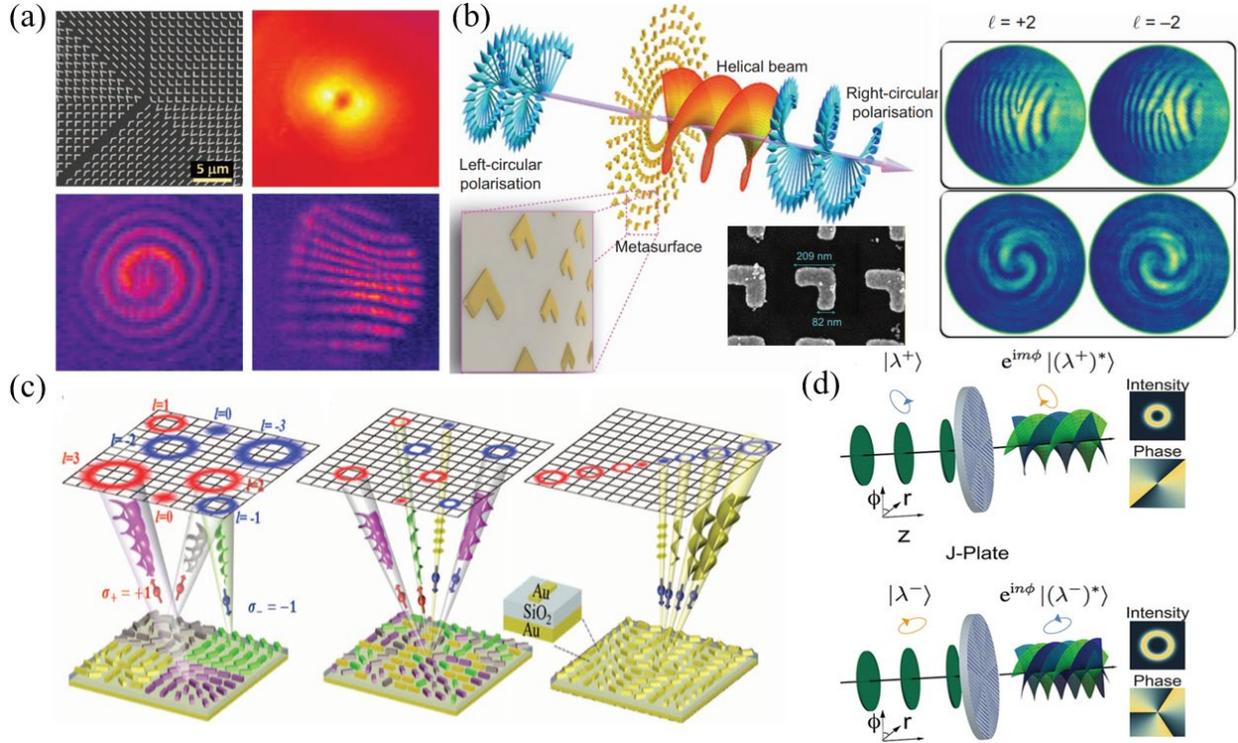

**Fig. 2. Nanophotonic generation of optical vortices using metasurfaces.** (**a**) A dynamic phase metasurface imprinted with a spiral phase profile. (**b**) A spin-to-orbital conversion L-shape metasurface for the generation of optical vortices with $l = \pm 2$. (**c**) An interleaved geometric metasurface for the simultaneous generation of multiple optical vortices carrying different OAM modes. (**d**) A J-plate metasurface based on a new concept of arbitrary spin-orbit conversion. The images in (a-d) were adapted from Refs. 36-39, respectively.

Metasurfaces consist of an array of subwavelength structures that can manipulate the wavefront of light in an almost arbitrary way. Metallic metasurfaces based on a resonant phase response from subwavelength scattering structures have been developed to generalize the interfacial reflection and refraction laws of light[36, 37]. By tailoring the shape, size, and material composition of a meta-atom (single pixel of a metasurface), a dynamic phase response covering the 2π phase change can be achieved. This has opened the possibility of using an ultrathin metasurface to imprint any desired phase profiles. As such, metasurface-based optical vortices generation has recently been realized. As an example, a V-shaped metasurface[36] was designed to pattern a spiral phase gradient, as shown in Fig. 2(a), leading to a metasurface-based spiral phase plate with high resolution. The experimentally obtained doughnut-shaped intensity distribution, as well as the interference pattern between an optical beam passing through the metasurface and a Gaussian reference beam, have confirmed the generation of an optical vortex beam with a topological charge of +1.

In addition to the dynamic phase response in plasmonic resonators, metallic metasurfaces have also been employed to implement the Pancharatnam–Berry geometric phase. The geometric phase offers an accurate phase manipulation approach through the spin-to-orbit conversion that couples an incident circularly polarized beam into an opposite helicity. Owing to the total angular momentum conservation, the incident SAM of a circularly polarized light changes its helicity after passing through the geometric metasurface, giving rise to an OAM state with $l = \pm 2$ in the cross-polarization state. As such, a plasmonic geometric metasurface[37] made of L-shaped plasmonic nanostructures was used to create an optical vortex beam with $l = \pm 2$, as shown in the Fig. 2(b). Here the geometric metasurface with strong circular birefringence acts as a spin-to-orbit angular momentum conversion plate. In general, one metasurface imprinted with a single spiral phase pattern can only create a given OAM state, limiting its application prospects. To further expand the access to more OAM states by a single metasurface, the shared-aperture antenna array initially developed for radar applications[51, 52] was introduced to the area of nanophotonics[53, 54]. The shared-aperture antenna concept is an effective design strategy by combing sets of antennas to achieve multiple functionalities, without increasing the size of the shared-aperture device. This concept was applied to the design of a geometric metasurface, providing a new route to create multifunctional metasurfaces. As an example, an interleaved geometric metasurface[38] capable of simultaneously generating multiple OAM states was introduced (see Fig. 2c), albeit it faces the speckle noise problem due to the multiplexing crosstalk.

Even though tremendous research progress has been made in plasmonic metasurfaces based on noble metals, their applications are often limited by strong material loss in the ultraviolet and visible spectral ranges due to the interband transitions. All-dielectric metasurfaces based on the principles of either geometric phase or Mie resonances have recently attracted strong interest, paving the way for developing efficient flat optics for optical vortex generation. In this context, the recent development of J-plates (J represents the total angular momentum) marks a milestone of using dielectric metasurfaces for optical vortex generation[39]. Unlike q-plate metasurfaces generating an optical vortex beam with only two possible OAM states of $l = \pm 2$, J-plate metasurfaces can imprint an arbitrary vortex phase to the coupled circular polarization states. The basic principle of a J-plate is based on the simultaneous modulation of both propagation phase and geometric phase responses of a meta-atom. This allows the geometric phase principle to be extended from circular polarization states to any orthogonal polarization states, and each state can carry an arbitrary OAM state. As an example, Figure 2(d) shows that two orthogonal elliptical polarization states of an incident optical beam can carry different OAM states with $l = +2$ and +3 after passing through a J-plate metasurface.

In addition to metasurfaces, soft materials have also been used for the generation of optical vortices. In this context, cholesteric liquid crystals (ChLCs) are liquid crystalline phases in which the director of rod-like molecules self-assembles into a helical structure, forming a one-dimensional soft photonic crystal[55]. For light propagating along the helix axis, they exhibit a Bragg reflection band for circularly polarized light with the same handedness as the helix over the wavelength range $n_o p - n_e p$, where $n_o$ and $n_e$ are ordinary and extraordinary refractive indices and $p$ is the helix pitch[56]. Recently, spatial engineering of the reflective geometric phase has been reported by controlling the initial alignment of planar ChLCs[57, 58], leading to the demonstration of various optical elements including optical vortices[59].

Notably, the intrinsic stimuli-responsive characteristic of the soft matter material (e.g., ChLCs) will endow corresponding elements with the dynamic tunability. This feature makes such self-organized ChLCs a promising platform for active flat optics, although the performance of their devices is limited by their large pixel sizes on the order of several micrometers. A recent approach combines nanostructured metasurfaces with liquid crystals, demonstrating active control over light beams by high-resolution metasurfaces for dynamic beam steering with large angles[60]. Therefore, the integration of metasurfaces with tunable material platforms, including liquid crystals[61-63], phase-change materials[64-66], thermo-responsive polymers[67] could provide opportunities to develop the next generation of augmented reality, light detection and ranging (LIDAR), holographic display technologies, and artificial intelligence.

## 3.2. Resonators-based generation of optical vortices

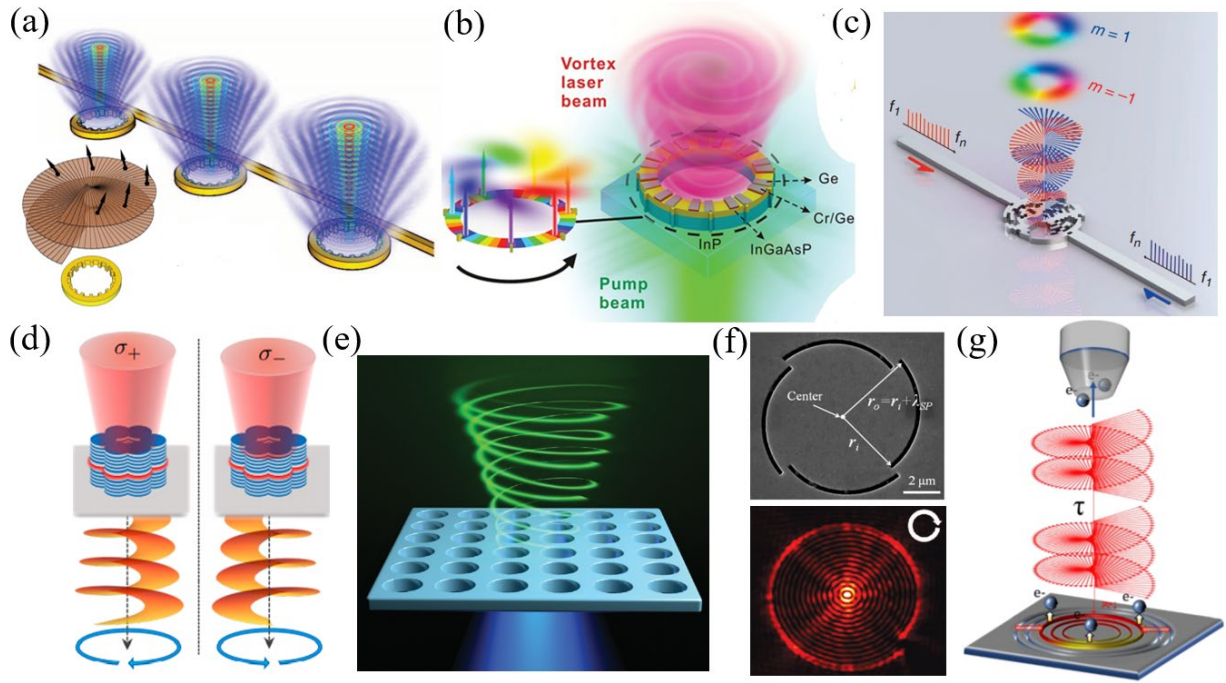

**Fig. 3. Generation of OAM-carrying optical vortices via dielectric and plasmonic resonators.** (**a**) An array of OAM vortex WGM emitters. (**b**) An OAM microlaser. (**c**) A silicon waveguide-based OAM multiplexer. (**d**) Principle of a chirality-tunable OAM microlaser. (**e**) A perovskite BIC metasurface. (**f**) A plasmonic vortex resonator. (**g**) Schematic of a plasmonic skyrmion imaged through a time-resolved two-photon photoemission electron microscope. The images in (a-g) were adapted from Refs.41-47, respectively.

Dielectric and plasmonic resonators with strong mode coupling offer another viable approach in nanophotonics to creating different optical vortices. A single resonator can strongly interact with electromagnetic waves through the constructive interference of coupled modes reflected from its boundaries. To achieve high quality-factor resonances, one can improve reflectivity through the total internal reflection at glancing angles of incidence in whispering-gallery-mode (WGM) resonators[41, 43]. Alternatively, destructive interference of strongly coupled modes has also been exploited, which are connected to the bound states in the continuum (BIC)[45, 68-70]. Recently, both the WGM and BIC resonators have been implemented for the generation of optical vortices.

More specifically, a silicon-integrated WGM resonator was designed to generate different OAM modes in the near-infrared wavelength regime[41]. The WGM resonator has a ring diameter of 3.9 µm, embedded with subwavelength angular grating structures along the azimuthal direction to couple the confined high-quality-factor resonant mode carrying the OAM into free space (Fig. 3a). The outcoupled optical vortex mode can be determined from the periodic angular gratings, leading to the generation of eight different OAM beams with $-4 < l < +4$. Since the WGM features a very large quality factor ($\sim 10^8$)[71], it can be used as a laser cavity. Recently, a WGM-based vortex microlaser[43] capable of producing an OAM-carrying vortex lasing mode was demonstrated (Fig. 3b). This WGM-based microlaser has a periodic alternation of Ge and Cr/Ge nanostructures on top of a 500 nm-thick InGaAsP multiple quantum well substrate. Like the case of a WGM vortex emitter, the emitted OAM lasing mode can be modulated by periodic Ge and Cr/Ge nanostructures.

Although WGM resonators offer compact generation of optical vortices, their inherent narrow bandwidths prevent them from being compatible with wavelength-division multiplexing in optical communications. A silicon-based OAM multiplexer chip with a broad bandwidth ranging from 1450 to 1650 nm was demonstrated (see Fig.3c)[42]. The silicon device has a small footprint and contains phase-arrayed silicon nanostructures that can generate OAM modes based on the excitation from single-mode waveguides. The phase modulation originates from both the propagation phase delay caused by optical path and the resonance phase delay from local nanostructures. Broadband waveguide signals can be used to couple into the OAM multiplexer chip and thereafter generate vertically emitting OAM modes of -1 and +1 controlled by the waveguides on the left and right sides, respectively. In addition to the vortex emitters mentioned above, dielectric vortex emitters can also be engineered to manipulate optical chirality. Recently, an integrated OAM microlaser[44] was developed to control the chirality of the emitted helical wavefront. This chirality-tunable OAM microlaser is composed of six micropillars embedded with single InGaAs quantum wells. The spin–orbit coupling originating from photon hopping between neighbouring micropillars leads to the chirality control of emitted photons, which can be optically controlled by pumping the microlaser with different circular polarization states (see Fig. 3d).

In addition, 2D plasmonic crystals were designed to observe polarization vortices in momentum space[69, 70]. The nodal intersection caused by the symmetry of designed photonic crystal slabs can impart a vortex in the polarization vector field, which is fundamentally governed by the BIC.[72] As an example, optical vortices with topological charges of $\pm 1$ and −3 were observed in a square lattice, and −2 and $\pm 1$ in a hexagonal lattice, respectively.[70] Apart from this, a dielectric perovskite vortex laser that supports bound states in the continuum (BIC) has also been demonstrated[45]. The BIC-based vortex laser is made of a 220 nm lead bromide perovskite film that is patterned with a square array of circular holes (see Fig. 3e). This result provides a viable approach to achieving ultrafast switching of vortex lasing modes within only 1 to 1.5 picoseconds. On the other hand, plasmonics opens the possibility of producing subwavelength plasmonic vortices carrying different OAM modes [46]. Plasmonics provides a new path to generate optical vortices in the near-field. As an example, Figure 3f presents a plasmonic vortex lens[46] consisted of sectional Archimedean spiral grooves in a gold film, which can be used to observe plasmonic vortices. Through precisely controlling the number of grooves, as well as the distance between inner and outer radii of each groove, different plasmonic vortices carrying distinctive OAM modes can be generated (see Fig. 3f). In addition to plasmonic vortex fields, skyrmion-like spin textures generated by plasmonic fields have also been designed to enhance quasiparticle interactions, such as plasmon-exciton and plasmon-phonon couplings [47, 73]. As such, plasmonic skyrmions were recently applied to observe more advanced light-matter interactions[74]. New experimental techniques, including time-resolved two-photon photoemission electron microscopy[47, 48, 75] and phase-resolved near-field optical microscopy[76], were recently developed to dynamically observe the formation and propagation of plasmonic vortex fields[77, 78] and optical skyrmions[79].

## 4. Nanophotonic detection of optical vortices

Another key aspect of OAM research and applications is the detection of optical vortices using nanophotonic devices. Conventional vortex beam detection commonly relies on the interference of a helical wavefront with a reference beam[33], or on the mode conversion of a helical wavefront into a fundamental plane wave via spatial light modulators[80, 81] and passive diffractive optical elements[82, 83]. However, these methods inevitably use bulky free-space optics to distinguish and separate the vortex beams, making their associated devices difficult for integration. Recently, nanophotonic research has opened an unpreceded opportunity to use miniaturized photonic devices for the OAM detection. Here, we provide a mini-review on recent development of on-chip optical vortices detectors, ranging from plasmonics[23, 24, 49] to photocurrent detectors[50].

### 4.1. Plasmonic detection of optical vortices

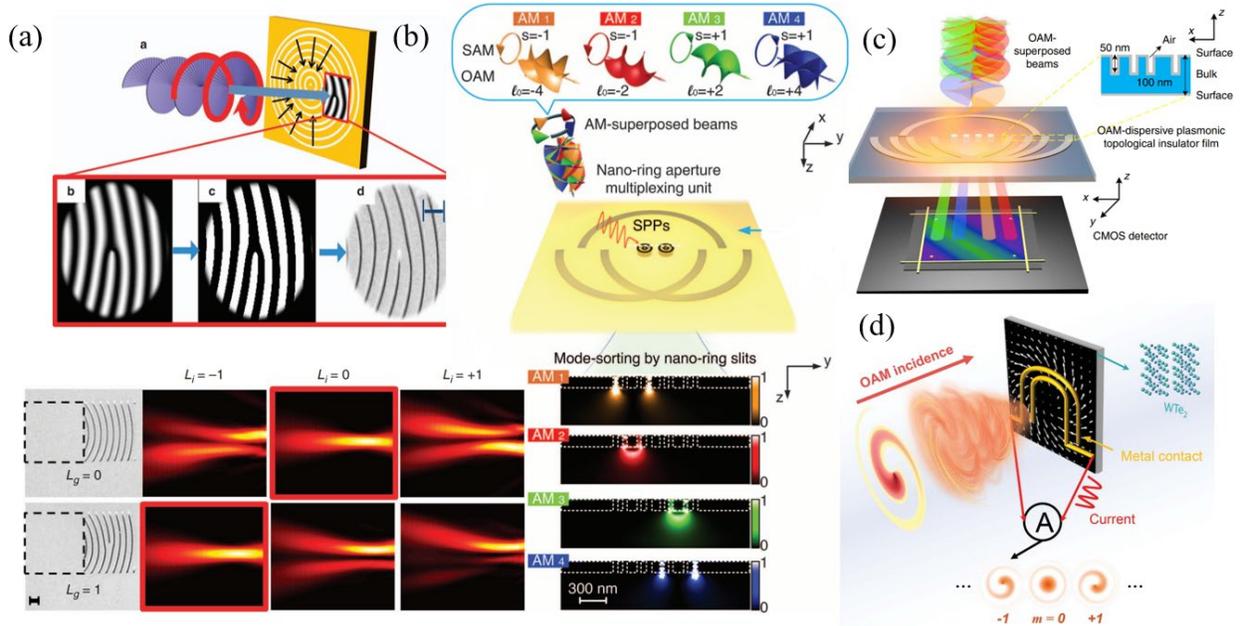

**Fig. 4. Plasmonic and photocurrent detection of optical vortices.** (**a**) A holographic plasmonic metasurface was used to detect a single OAM mode in the near-field. (**b**) An ultracompact OAM-demultiplexing chip. (**c**) A plasmonic topological insulator-based OAM nanometrology chip. (**d**) Photocurrent detection of incident optical vortices. The images in (a-d) were adapted from Refs. 49,23,24,50 respectively.

A plasmonic nanogroove pattern was designed to detect a single OAM mode, based on the principle of near-field holography, through the interference of focused surface plasmon polaritons and an incident optical vortex field. In the detection process, an incident optical vortex beam with a given helical phase is converted into a focused plasmonic field, under which a subwavelength grating array was engraved to transmit the focused plasmonic field into a photodetector (see Fig. 4a)[49]. On the other hand, incident optical vortices with different OAM modes cannot be coupled into a focused plasmonic field, which are blind to the photodetector.

Recently, a new concept of on-chip noninterference OAM multiplexing of broadband light was introduced (see Fig. 4b)[23]. Here, coaxially-superposed OAM carrying beams, with four selected angular momentum modes of $l=-4$, $s=-1$ (AM$_1$), $l=-2$, $s=-1$ (AM$_2$), $l=+2$, $s=+1$ (AM$_3$), and $l=+4$, $s=+1$ ($AM_4$), propagate through a nanoring aperture multiplexing unit consisting of shallow nano-grooves and the spatially-shifted mode-sorting nano-ring slits in a gold thin film. The nano-groove structures were used to convert the AM modes carried by photons into plasmonic fields and to spatially route the excited plasmonic AM modes to the locations of nano-ring slits. The incident OAM-carrying beams can excite four different plasmonic AM modes with a distinguished spatial separability, based on the conservation of total angular momentum. The formation of the spatial separability by nano-grooves paves the way for plasmonic AM mode sorting. Owing to the distinctive AM mode-sorting sensitivity by nano-ring slits, the plasmonic AM modes can be selectively coupled out through the slits of different sizes and spatial shifts. Furthermore, the non-resonant AM mode-sorting sensitivity by the nano-ring slits enables AM multiplexing over a broad bandwidth.

Tremendous progress has been made in understanding the fundamental physics of the combined fields of plasmonics and metamaterials and in utilizing them for a great variety of applications. However, the scope of plasmonic metamaterials is often limited by energy dissipation of noble metals in the UV to visible spectral range. Traditional noble metals such as gold and silver suffer from high losses due to interband transitions[84-86]. They are furthermore CMOS-incompatible and show low thermal stability. Topological Insulators[87-89] have unique highly conducting symmetry-protected surface states while the bulk is insulating, making them attractive for various applications in condensed matter physics. Recently, topological insulator materials have been applied for both near-[90] and far-field[24] wavefront manipulation of electromagnetic waves yielding superior plasmonic properties in the ultraviolet to visible wavelength range. In this context, an entirely new concept of CMOS-integratable OAM nanometrology in an ultrathin topological insulator film was demonstrated (see Fig. 4c)[24]. A dispersion analysis suggests that the surface plasmon polaritons in the topological insulator $Sb_2Te_3$ thin film exhibit superior plasmonic figures of merit as compared to the noble metals (e.g., gold). Applying the superior plasmonic effect in an ultrathin $Sb_2Te_3$ film with a thickness of 100 nm to the linear displacement engineering of plasmonic AM modes at a nanoscale, a CMOS-integratable OAM nanometrology device with a low modal crosstalk of less than -20 dB was successfully demonstrated. The demonstrated ultrathin OAM nanometrology device with a small footprint of 13.4 µm by 10.6 µm opens exciting avenues for the development of ultrathin and CMOS-compatible OAM detectors.

### 4.2. Photocurrent detection of optical vortices

OAM detection commonly relies on the design of structures to offer OAM selectivity that can separate different helical wavefronts in space, since photodetectors are only sensitivity to the intensity of light. Recently, a novel concept of OAM-dependent photocurrent detection has been introduced (see Fig. 4d), providing a new platform for the OAM detection[50]. This concept is based on the OAM-dependent photo-galvanic effect, that is, the generated photocurrent is proportional to the helical phase gradient. In this work, OAM-carrying incident beams were modulated into left- and right-handed circular polarization states, showing distinctive photocurrent responses that are proportional to the quantized OAM mode number. In order to measure the photocurrent, the electrodes were carefully designed into a U shape on tungsten ditelluride substrate. Light beams with OAM order from -4 to +4 were successfully distinguished by this OAM photodetector, providing a new route to develop a compact OAM detector.

## 5. Applications of optical vortices

Since the pioneering work by Allen et al. on the OAM in an optical vortex beam, twisted light has been applied for many photonic applications, including optical trapping[27, 91-96], quantum entanglement[97-100], optical communications[101-104], superresolution[105-111], phase-contrast imaging[112-114], pattern recognition[115, 116], and recently optical holography[117-120]. Here we particularly focus on the applications of optical vortices in optical trapping[27, 95, 96], phase-contrast imaging[121], and holography[122, 123].

### 5.1. Optical vortices for advanced trapping and imaging

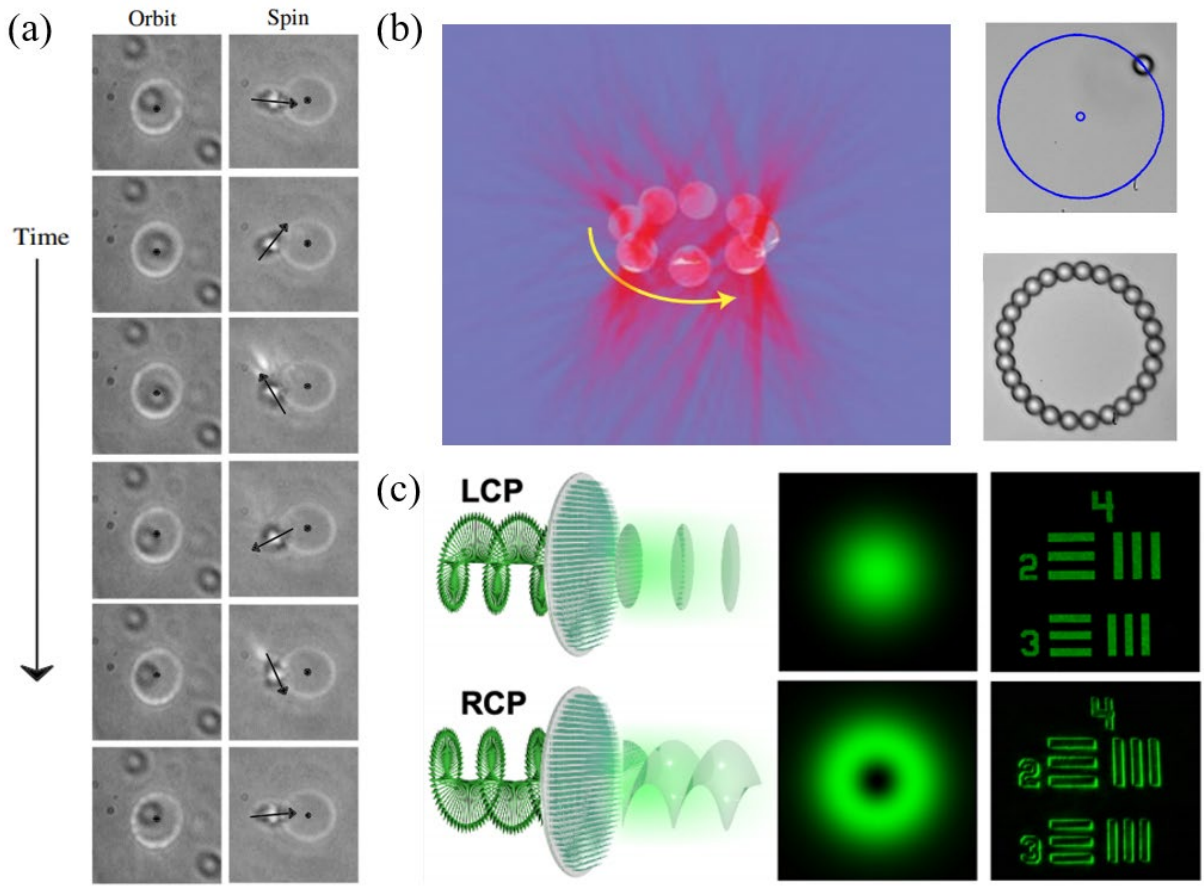

**Fig. 5. Optical trapping and edge-enhancement imaging using optical vortices.** (**a**) A particle trapped by the OAM and SAM of light, respectively. (**b**) A particle neckless consisting of 26

particles, trapped by a vortex beam. (**c**) Edge-enhancement imaging based on an optical vortex beam. The images in (a-c) were adapted from Refs 27, 95 and 121, respectively.

Optical trapping, also known as optical tweezers, is the Nobel Prize-winning technology[124]. Optical tweezers can grab particles, atoms, molecules, and living cells with the use of a focused laser beam. For typical optical tweezers, a tightly focused Gaussian beam creates a bright spot to push small particles towards the center of the beam and to hold them there. In 1995, Rubinsztien Dunlop et al. experimentally confirmed that an optical vortex trapping carrying the OAM can optically induce a rotation force to a particle[92]. Since then, significant progress has been achieved in the field of twisted light optical trapping[27, 94]. OAM traps have been confirmed not only to rotate particles with different speeds[125] or stop the rotation[126], but also to trap multiple particles[127-129].

Early work[91-93] in optical tweezers focused on the trapping of particles at or near the optical axis. OAM traps open the possibility of control over small particles away from the optical axis, through converting the OAM in an optical vortex beam into a mechanical force that can move the particle around. Depending on the SAM and OAM components in an optical vortex trap, a particle can be rotated either around the its own axis or a circular orbit, respectively (see Fig. 5a)[27]. In addition, an optical vortex trap can simultaneously trap multiple particles together in an organized fashion. As shown in Fig. 5b(right), a particle necklace composed of 26 polymer particles can be simultaneously trapped by a single optical vortex beam[95]. This necklace of particles was rotated by a vortex trap, with the rotation speed dictated by the gradient and scattering forces resulting from a vortex beam.

In additional to the application of optical trapping, optical vortex beams have also been used for improving image quality[130-132]. Based on the imaging filtering by a 4f system, the image of a planar object is formed by a convolution of the object transmission function with the point-spread function of a spatial filter[133]. When a vortex-phase-mask is used in the Fourier imaging system, it has a capacity to generate edge-enhanced images. Specifically, integration of the intensity-uniform area leads to destructive interference and a dark background, due to the phase difference of π in the opposite azimuth of the vortex phase. But at the edges of structure where the object has strong gradients in amplitude and phase retardation, it cancels the destructive interference and the signal is enhanced[134-136]. A recent example presents the use of a spin-multiplexing metasurface for switchable diffraction-limited and spiral-phase-contrast imaging (see Fig. 5c)[121]. The designed dielectric metasurface can imprint an OAM vortex phase to an incident right-handed circular polarization, while keeping a plane wave output when the incident beam is left-handed circular polarization. Therefore, by controlling the incident polarization, diffraction-limited and edge-enhancement imaging processes can be selectively obtained. Notably, unlike the polarization-dependent edge enhancement that is effective only in one direction determined from the incident polarization, OAM-based edge enhancement is isotropic. This work based on an ultrathin metasurface further extends the application of OAM for edge detection and advanced image processing.

## 5.2. Twisted light holography

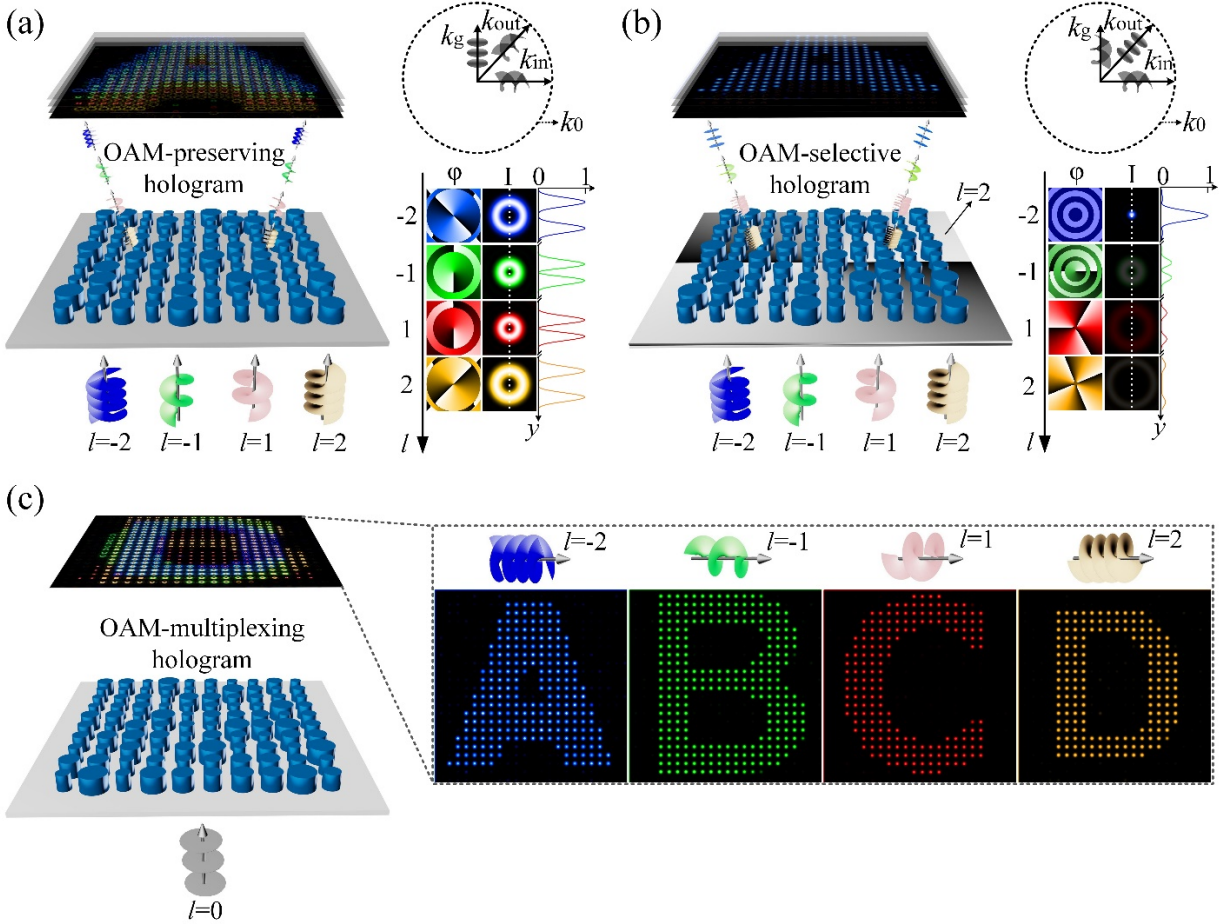

**Fig. 6. Principle of OAM holography.** (**a**) Schematic of an OAM-preserving hologram capable of transferring the OAM property from an OAM incident beam to a holographic image. The inset (right top) shows the OAM property transfer in the spatial frequency domain (k-space). The inset (right bottom) presents the phase ($\varphi$) and intensity ($I$) distributions of single pixels in the reconstructed holographic images, respectively. Pseudo colors are used to visualize different OAM modes. (**b**) Schematic of an OAM-selective hologram sensitive to a given OAM mode. The inset (right top) shows the OAM conversion from an incident OAM beam to a fundamental spatial mode after passing through an OAM-selective hologram. The inset (right bottom) presents the phase ($\varphi$) and intensity ($I$) distributions of single pixels in the reconstructed holographic images using different OAM modes. High intensity is achieved in each pixel whenever the incident light has a topological charge matching the design of the interface. (**c**) Schematic of an OAM-multiplexing hologram capable of reconstructing multiple distinctive OAM-dependent holographic images. Fig. 6 is adapted from Ref. 122.

Holography offers a vital platform for 3D light-field displays, LiDAR, optical encryption, and artificial intelligence, allowing read-out of image information at the speed of light. Recent advances in metasurface technology have allowed the use of ultrathin optical elements to achieve complete control of the amplitude, phase, and polarization of light, leading to a more versatile platform for digitizing optical holograms with nanoscale resolution. To increase a metasurface hologram bandwidth, essential for optically addressable holographic video displays with potentially ultrafast switching of image frames, different degrees of freedom of light including polarization[137], wavelength[138], and incident angle[139] have been exploited to carry and multiplex different holographic images. However, the bandwidth of a metasurface hologram has remained too low for any practical use.

Recently, the concept of twisted light holography was introduced, opening the possibility of storing holographic information in different OAM modes[140-143]. Due to the fact that the OAM degree of freedom of light has a theoretically unbounded set of orthogonal helical modes, OAM holography using incident OAM beams as independent information carriers holds great promise for largely improving the bandwidth of a single hologram. Unfortunately, due to the lack of OAM sensitivity, a conventional digital hologram fails to offer distinct responses to incident OAM modes. To demonstrate metasurface OAM holography principle, three types of meta-holograms[122] with discrete spatial frequency distributions are designed, including OAM-conserving (Fig. 6a), -selective (Fig. 6b), and -multiplexing (Fig. 6c) meta-holograms, respectively. To preserve the OAM property in each pixel of a reconstructed holographic image, it is necessary to spatially sample the holographic image by an OAM-dependent two-dimensional Dirac comb function to avoid spatial overlap of the helical wavefront kernel, i.e. creating OAM-pixelated images. In this context, the constituent spatial frequencies ($k_g$ in the momentum space) of an OAM-conserving meta-hologram add a linear spatial frequency shift to an incident OAM beam ($k_{in}$). As such, outgoing spatial frequencies leaving the meta-hologram ($k_{out}$) possess a helical wavefront inherited from the incident OAM beam, which implies that the OAM-conserving meta-hologram could create OAM-pixelated holographic images.

Mathematically, adding a spiral phase plate that features a phase distribution of $l\varphi$ ($l$ and $\varphi$ refer to the topological charge and the azimuthal angle of a phase change, respectively) on an OAM-conserving meta-hologram leads to an OAM-selective meta-hologram, of which the constituent spatial frequencies ($k_g$) carry a helical wavefront (Fig. 6b). In this case, owing to the OAM conservation, only a given OAM mode with an inverse topological charge (-$l$) can recover the fundamental spatial mode with a relatively stronger intensity distribution in each pixel of the holographic image, and hence to distinctively reconstruct the holographic image. Consequently, the OAM selectivity discussed above can be further extended to realize an OAM-multiplexing meta-hologram by superposing multiple OAM-selective meta-holograms to reconstruct a range of OAM-dependent holographic images (Fig. 6c). The latter demonstration suggests that different OAM modes can be adopted to carry independent information channels for holographic optical multiplexing. As an example, we show that incident OAM beams with topological charges of $l$ = −2, −1, +1, and +2 can independently reconstruct distinctive holographic images (alphabet letters of A, B, C, and D) from a single OAM-multiplexing meta-hologram, respectively.

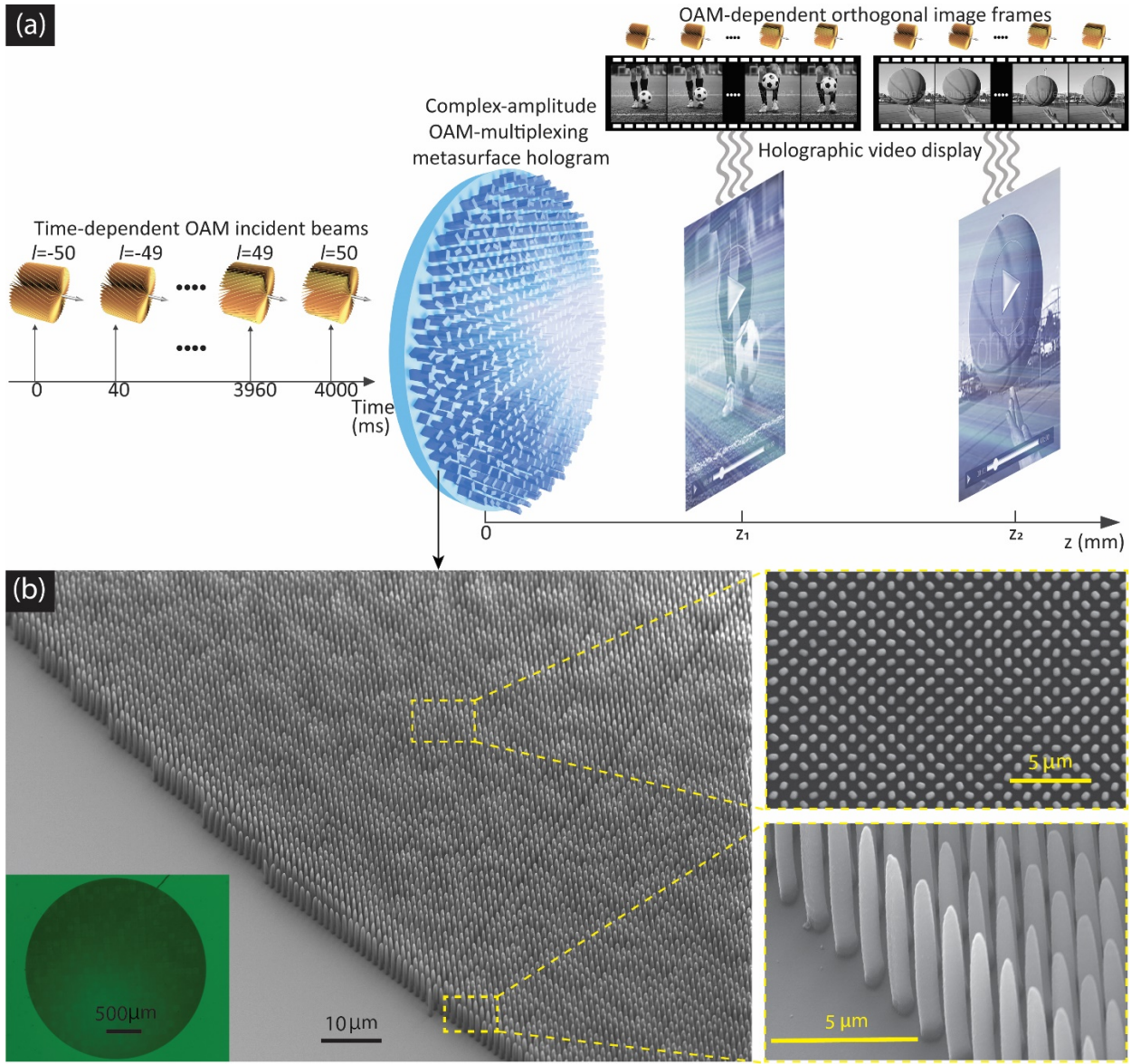

**Fig. 7. Principle of high-bandwidth twisted light holography based on a complex-amplitude metasurface hologram.** (**a**) Time-dependent OAM incident beams impinge on a large-scale complex-amplitude metasurface hologram to reconstruct two holographic videos consisting of a large number of OAM-dependent image frames. (**b**) SEM and optical (bottom left inset) images of a fabricated complex-amplitude OAM-multiplexing metasurface hologram. Enlarged SEM images of nanopillars are given on the right side. Fig. 7 was adapted from Ref. 123.

In 2020, Ren et al. presented the design and 3D laser manufacturing of a large-scale complex-amplitude metasurface for ultrahigh-dimensional OAM-multiplexing holography in momentum space[123]. Such high-bandwidth metasurface holograms allow us to demonstrate lensless reconstruction of up to 200 of orthogonal image channels encoded within an OAM signature, leading to optically-addressable holographic video displays. Without loss of generality, we selected OAM modes with helical mode indices ranging from -50 to +50 to be sequentially incident on a large-scale complex-amplitude OAM-multiplexing metasurface hologram for addressing OAM-dependent orthogonal image frames (Fig. 7a), with two distinct holographic videos being simultaneously reconstructed. The independent reconstruction of holographic videos in two different planes suggests that our a pproach can be applied for 3D holography. Unlike the conventional planar metasurfaces with restricted degrees of freedom in a 2D plane, we introduce the design and laser-based printing of 3D meta-optics, in which the height ($H$) and in-plane rotation ($\vartheta$) of a birefringent polymer nanopillar are employed to independently control the amplitude and phase responses of transmitted light, respectively (Fig. 7b).

## 6. Summary

Recent advances in nanotechnology have been a major propellant of using miniaturized nanophotonic devices for the generation and detection of optical vortices, opening new frontiers in wide-ranging photonic applications. In this article, we provided a mini review on recently developed nanophotonic devices for the generation and detection of different optical vortices. The highlighted nanophotonic devices by this article, however, take up a small portion of cutting-edge nanophotonic devices developed for twisted light manipulation. Indeed, the field of nanophotonic manipulation of twisted light has undergone rapid growth in the last decade, generating a large volume of milestone works that cannot be fully covered by this paper. Readers having more broad research interests in the field can benefit from other review articles[144, 145].

We have also highlighted some of applications of twisted light, focusing on optical trapping, phase-contrast imaging, and optical holography. Our recently introduced OAM holography concept has the potential to significantly boost the image storage capacity of a single passive hologram, which may greatly improve the future information systems. We believe nanophotonic manipulation of twisted light will continue to make strong impact not only on fundamental understanding of light-matter interactions that harness the OAM, but also on wide-ranging applications that demand ultracompact, ultrahigh-capacity, and ultrahigh-speed optical devices and systems.

**Acknowledgement**

C.L. acknowledges the scholarship support from the China Scholarship Council. H. R. acknowledges the funding support from the Macquarie University Research Fellowship (MQRF) from Macquarie University. S. A. M. acknowledges the funding support from the Deutsche Forschungsgemeinschaft, the EPSRC (EP/M000044/1), and the Lee-Lucas Chair in Physics.